\documentclass[bookmarks=true, pdfnewwindow=true, colorlinks=true, linkcolor=xlinkcolor, citecolor=xlinkcolor, filecolor=xlinkcolor, urlcolor=xlinkcolor, final=true,%
a4paper,
prd,
twocolumn,
superscriptaddress,
nofootinbib,
nobibnotes,
amsmath,amssymb,
aps,
fleqn,
floatfix
]{revtex4-2}
\usepackage{amsfonts,graphics,color}
\usepackage[english]{babel}
\usepackage{graphicx}
\usepackage{subfigure}
\usepackage{amsmath}
\usepackage{amssymb}
\usepackage{dcolumn}
\usepackage{comment}
\usepackage{multirow}
\usepackage{xcolor}
\usepackage{placeins}
\usepackage{caption}
\usepackage{orcidlink}
\definecolor{xlinkcolor}{cmyk}{1,1,0,0}
\usepackage[normalem]{ulem}
\definecolor{myblue}{rgb}{0.05,0.1,0.5}

\begin{document}

\preprint{INR-TH-2026-012}

\title[s]{Dark-matter relay for ultra-high-energy cosmic rays}

\author{Mikhail Sekretov\,\orcidlink{0009-0007-9145-8429}}
\thanks{E-mail: sekretov.mk21@physics.msu.ru}
\affiliation{Chair of Particle Physics and Cosmology, Physics Department,  Lomonosov Moscow State University, Moscow 119991, Russia}
\affiliation{Institute for Nuclear Research of the Russian Academy of Sciences, Moscow 117312, Russia}

\begin{abstract}
Astrophysical sources of ultra-high-energy cosmic rays (UHECRs) remain unidentified, and no nearby cosmic accelerators are established even for the highest-energy events. In addition, puzzling correlations of a fraction of UHECR flux with distant blazars were found by HiRes and, recently, Telescope Array experiments. Here we propose that dark matter can provide a mechanism to effectively transfer UHECRs from distant sources: dark-matter particles are boosted by UHECRs near the source and pass their momentum to baryons close to the observer. The probability of this relay is not negligible because of the growth of dark-matter–baryon cross section with energy. We demonstrate that the effect is observable for the fermionic dark matter of the mass $m_\mathrm{DM} \sim 25 \, \mathrm{MeV}$ and discuss possible observational signatures of this mechanism.

\end{abstract}
\maketitle

\section{Introduction}
\label{sec:intro}
At present, the origin of ultra-high-energy ($\lesssim 10^{19} \,\mathrm{eV}$) cosmic rays remains unknown: no suitable astrophysical accelerators are found in the local Universe, while cosmic rays with such extreme energies are unable to propagate over cosmological distances \cite{Greisen:1966, Zatsepin:1966}. For instance, the extremely energetic “Amaterasu” particle, recently observed by the Telescope Array experiment \cite{TelescopeArray:2023Amaterasu}, spectacularly points back to a large-scale cosmic void.

Active galactic nuclei are promising sites of cosmic-ray acceleration. Two decades ago, correlations were reported between cosmic-ray arrival directions  detected by the High Resolution Fly’s Eye (HiRes) experiment and BL Lacertae (BL Lac) objects \cite{Gorbunov:2004bs}. These results were later confirmed using a larger sample of cosmic-ray events \cite{HiRes:2005kxa}, and similar correlations have recently been identified in the Telescope Array data \cite{Kudenko:20256P}, albeit with a smaller significance.
However, these BL Lac objects are located at cosmological distances, and charged nucleons emitted from active galactic nuclei are expected to be deflected by cosmic magnetic fields. In contrast, the angular separations found in these correlations are significantly smaller than the expected deflections. Only neutral particles could travel along straight lines for such distances, and no compelling options exist within the Standard Model \cite{Tinyakov:2006jm}. This discrepancy calls for a new physical explanation. 

Since dark-matter particles are generally assumed to be stable and neutral, explanations for the aforementioned correlations—and possibly other, similar effects—may be sought by investigating the interaction of cosmic rays with dark matter. In this work, we propose a mechanism in which dark matter acts as a “relay” for cosmic rays: in the first stage, a dark-matter particle is accelerated to high energies by a relativistic nucleon, and subsequently transfers this energy back to baryonic matter, potentially traversing cosmological distances in the process. 

In Section \ref{sec:2}, we estimate the cross section required for an accelerated nucleon to collide with a dark matter particle. Using the typical energy and mass dependence of the cross section, we determine the necessary parameter values and then evaluate the probability of such a collision. In Section \ref{sec:3}, we estimate the probability of a subsequent collision with stationary baryonic matter, relatively close to the observer. In Section \ref{sec:4}, we present an example of a particular model of dark matter in which the required cross sections are reached. We briefly conclude in Section \ref{sec:5}.

\section{Estimate of the Cross Section and Interaction Probability}
\label{sec:2}
In order to estimate the cross-section values for which the probability of interaction between accelerated nuclei and dark matter is sizable, we express the cross section as  
\begin{equation}
    \sigma_c = \frac{1}{n  \lambda} = \frac{m_\mathrm{DM}}{\rho  \lambda}.
\end{equation}
Here $n$, $\rho$ and $m_\mathrm{DM}$ denote the number density, mass density, and mass of dark-matter particle, respectively, while $\lambda$ is the mean free path of the accelerated nucleus in the dark-matter medium. We will be primarily interested in the ratio of the cross section to the mass, so the number density is expressed in terms of the mass and density. As a result, the cross section is, a linear function of the dark-matter particle mass.

For the first order-of-magnitude estimate, we set $\lambda$ equal to the size of the region with dark-matter density $\rho$. 

Although an interaction between an accelerated nucleon and dark matter is more likely to occur in the central region of a galaxy, where the density is significantly higher, for completeness we consider three characteristic scales of $\lambda$: the galactic nucleus, the galactic scale, and the scale of galaxy clusters.
\begin{figure}[t]
    \includegraphics[width=\columnwidth]{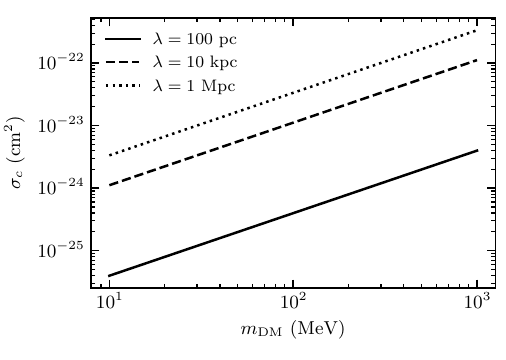}
    \caption{Estimates of the required cross section $\sigma_c$ versus mass
$m_\mathrm{DM}$ calculated from Eq.~(1). Three different spatial
scales $\lambda$ are considered.}
    \label{fig:cross_section_lambda}
\end{figure}
\newpage

\begin{center}
	{Table 1. Interaction scales}
\end{center}
\begin{center}	
\begin{tabular}{c c c c}
	\hline
	\hline
	Scale & $\lambda$, kpc & $\rho$, GeV/cm$^3$  \\
	\hline
	Core & 0.1 & 840   \\
	
	Galaxy & 10 & 0.3  \\
	
	Galaxy cluster & $10^3$ & $10^{-3}$   \\
	
	\hline
\end{tabular} \\
\end{center}

Table 1 summarizes the estimates of the dark-matter density at different scales. For the core scale, we adopt the value from \cite{Sofue2020}, although the estimate there was obtained for the Milky Way. The value at the galactic scale is widely quoted in the literature, for example in \cite{GorbunovRubakov2011}. It is generally assumed that the dark matter density in galaxy clusters exceeds the cosmic mean by several hundreds or even thousands \cite{GorbunovRubakov2011}; therefore, for the purpose of our estimate we take $\sim 10^3\cdot\rho_\mathrm{crit} \sim 10^{-3}$ GeV/cm$^3$.

The required cross-section values obtained in this way look large, given that dark matter interacts very weakly (Fig.~1). Nevertheless, as we will demonstrate in Section \ref{sec:4}, in certain models the asymptotic behavior of the cross section is such that it can reach values on the order of $10^{-25}$ - $10^{-24} \, \mathrm{cm} ^2$ at high energies, while remaining consistent with existing constraints at low energies. In this context, the mass is constrained to a relatively narrow range around $25 \,\mathrm{MeV}$, assuming that the interaction occurs on a scale of 100 pc (cf. point A in Fig.~2(b)). For subsequent estimates, we will adopt these values for the cross section ($\sigma \sim \sigma_c = 10^{-25} \,\rm cm^2$) and mass.

We proceed to calculate the probability of a high-energy proton interacting with a dark-matter particle. For this purpose, we consider a spherical halo of radius $R = 200 \, \mathrm{kpc}$, in which dark matter is distributed according to the Burkert profile \cite{burkert1995structure},
\begin{equation}
	\rho(r) \equiv \rho_\mathrm{Bk}(r) = \frac{\rho_\mathrm{b}}{(1+r/r_{\rm b})\left(1 + \left(r/r_{\rm b}\right)^2\right)}, 
\end{equation}
where $\rho$ is the density of dark matter and $r$ is the distance from the center of the galaxy.

Since galactic nuclei are considered potential cosmic-ray accelerators, we assume that an accelerated proton originates from the center of the galaxy. The interaction probability $\mathrm{d}P(r)$ in a spherical shell $(r;r+\mathrm{d}r)$ is given by:
\begin{align}
\mathrm{d}P(r) 
       = \frac{\sigma}{m_{\rm DM}}\, \rho(r)\, \mathrm{d}r \,.
\end{align}

Following \cite{Lin_2019}, we take $r_{\rm b} = 8$ kpc and determinate $\rho_{\rm b}$ from the normalization \cite{refId0}:
\begin{equation}
\begin{aligned}
 4\pi \int_{0}^{R} \rho(r)\, r^2 \,\mathrm{d}r = M \sim 10^{12} M_{\odot}.
\end{aligned}
\end{equation}
To obtain the total probability, we integrate over the entire path,
\begin{equation} 
	P = \int_{0}^{R}\frac{\sigma}{m_{\rm DM}}\rho(r)\,\mathrm{d}r
\end{equation}
In this way, for $m_\mathrm{DM} = 25\, \mathrm{MeV}$ and $\sigma = 10^{-25} \,\rm cm^2$, we obtain $P \sim 0.2$.

This estimate may be made more precise if we depart from the approximate estimate of the cross section. However, the parameter $r_{\rm b}$ strongly depends on the specific galactic model. A simultaneous increase of the cross section and decrease of $r_{\rm b}$ can lead to a considerable increase in the probability. A definitive conclusion requires a refinement of the estimates used. However, we see that transfer of a cosmic-ray momentum to a dark-matter particle may be possible for a non-negligible fraction of cosmic rays. 
\section{Interaction of Accelerated Dark Matter with Baryonic Matter}
\label{sec:3}
We now consider a scenario in which an accelerated dark-matter particle arrives from a distant source to our Galaxy and undergoes a secondary interaction with a proton. The gas in our Galaxy is distributed in the form of a thin disk with the density profile \cite{Li:2022}:
\begin{equation}
	\rho(R, z) = \frac{\Sigma_{\rm g}}{2z_{\rm gas}}\exp(-R/R_{\rm gas})\mathrm{sech}^2(z/z_{\rm gas}),
\end{equation}
where $\Sigma_{\rm g} = 71.1$ $M_{\odot}/$pc$^{2}$, $R_{\rm gas} =  4.8$ kpc and $z_{\rm gas} = 130$ pc. We assume that the process takes place in the vicinity of the Earth; therefore, we adopt the Galactocentric distance  $R_{\odot}= 8.15$ kpc.

We consider the case of a trajectory passing through the scale height of the disk. The probability of interaction within the layer is given similarly to Eq.(3) by
\begin{equation}
\begin{split}
\mathrm{d}P(z) = \frac{\sigma}{m_\mathrm{p}} \rho(z)\, \mathrm{d}z.
\end{split}
\end{equation}
To calculate the total probability, we integrate over the scale height of the disk, taking $h \sim$ 100 pc as the half-thickness
\begin{equation}
\begin{split}
P &= \int_{0}^{h} \frac{\sigma}{m_p} \rho(R = R_\odot, z)\, \mathrm{d}z \\
  &= \frac{\sigma}{m_\mathrm{p}} \frac{\Sigma_\mathrm{g}}{2 z_{\rm gas}} \exp(-R_\odot/R_{\rm gas}) 
     \int_{0}^{h} \mathrm{sech}^2(z/z_{\rm gas})\, \mathrm{d}z \\
  &\sim 5 \cdot 10^{-5},
\end{split}
\end{equation}
where we still take $\sigma = 10^{-25}\, \rm cm^2$, and $m_{\rm p}$ is the proton mass.

Another scenario is a particle passing through the center of the Galaxy. For simplicity, we set $z=0$ and integrate over $R$ from 0 to $R_{\odot}$:
\begin{equation}
\begin{split}
P &= \int_{0}^{R_\odot} \frac{\sigma}{m_{\rm p}} \rho(R, z = 0)\, \mathrm{d}R \\
  &= \frac{\sigma}{m_{\rm p}} \frac{\Sigma_{\rm g}}{2 z_{\rm gas}} 
     \int_{0}^{R_\odot} \exp(-R/R_{\rm gas})\, \mathrm{d}R \sim 0.01
\end{split}
\end{equation}
The resulting probability is significantly higher, indicating a clear directional anisotropy of effect.

One more scenario involves interactions within the filaments of the large-scale structure of the Universe. This possibility is of a particular interest, since in this case the expected signal would be more isotropic. In addition, refs. \cite{Troitsky:2021, Kudenko:2024} noted a possible association of BL Lac-correlated HiRes cosmic rays with the local-filament pattern. Taking into account that the gas number density in a filament is $n \sim 10^{-4} $ cm$^{-3}$ \cite{Martizzi2019} and the characteristic size of the structures is $\lambda \sim 30-100 \, \mathrm{Mpc}$ \cite{Tanimura2020}, we obtain the cross section,
\begin{equation}
	\sigma = \frac{1}{n \, \lambda} \sim 10^{-23} \,\, \mathrm{cm}^2.
\end{equation}	
In our analysis, we adopt the value $\sigma \sim 10^{-25}$ cm$^2$, assuming that the interaction occurs on a scale of 100 pc (point A in Fig.~2(b)). However, an alternative scenario is also possible. In this case, as we will show in Sec. \ref{sec:4}, the dark matter mass is reduced ($m_{\rm DM}\sim 10$ MeV), and the characteristic spatial scale at which the first interaction occurs is of order 1 Mpc (B on Fig.~2(b)). The cross section is expected to be slightly below $10^{-23} \, \rm cm^2$; however, it remains closer that of case A.

Note that a symmetric setup in which both interactions take place in a filament \cite{Troitsky:2021} is possible. Using an estimate for the filament density $\rho \sim 10^2 \, \rho_\mathrm{crit}$ \cite{Jauzac:2012} and a characteristic length scale $\lambda \sim 30$ Mpc, we find that the cross section is of order $10^{-24}$ cm$^2$ for a mass $\sim 10$ MeV, which is rather close to derived value. Therefore, this scenario also remains viable.
\section{High-Energy Asymptotics of the Dark Matter–Nucleon Interaction Cross Section}
\label{sec:4}
We now turn to the demonstration that the required cross section is viable.
To this end, we consider a dark-matter model in which the dark-matter particle is a Dirac fermion $\chi$ interacting with protons via dark photons $A'_{\mu}$. The corresponding interaction Lagrangian can be written as \cite{Su:2024}
\begin{equation}
		\mathcal{L} = g_{\chi}\bar{\chi}\gamma^{\mu}\chi A'_{\mu} + \sum_{q}g_q \bar{q}\gamma^{\mu}q A'_{\mu} \, ,
\end{equation}
where $q$ denotes quarks, $g_{\chi}$ is the dark matter–dark photon coupling constant (we fix $g_{\chi} = 0.1$), 
and $g_q = Q_q e \epsilon$ is the dark photon–quark coupling constant, where $Q_q$ depends on the quark charge and $\epsilon$ is the kinetic mixing parameter. 

For definiteness, we take benchmark value $m_{\chi} = 25~\mathrm{MeV}$ and $m_{A'} = 3 m_{\chi}$. For a dark photon of this mass, $\epsilon \lesssim 10^{-4}$ \cite{Essig:2013, Fabbrichesi:2021, Gninenko:2023}. In our calculations, we take $\epsilon = 10^{-5}$. In that follows, dark-matter particle will be identified with the fermion $\chi$; therefore, the quantities $m_{\chi}$ and $m_{\rm DM}$ will be used interchangeably.

We consider the process:
\begin{equation}
	\chi(k) + q(xp) \longrightarrow \chi(k') + q(p') 
\end{equation}
where $x$ denotes momentum fraction, $p$ and $p'$ is a four-momenta of the initial and final protons, $k$ and $k'$ those of the dark-matter particle. For the model we adopt, the differential cross section is expressed as \cite{Li:2025, Su:20241}:
\begin{equation}
\begin{split}
\frac{\mathrm{d}\sigma_{\rm \chi p}}{\mathrm{d}T_{\chi}} 
&= \frac{\lvert \vec{k'}\rvert}{16 \pi m_{\chi} Q^2 \sqrt{E^2_{\rm p} - m_{\rm p}^2}} \\
&\quad \times \sum_{\rm q = u, d} \int_{-1}^{\cos\theta_{\rm max}} 
f_{\rm q}(x, Q^2) \, \overline{\lvert \mathcal{M}(x)\rvert^2} \, \mathrm{d} \cos \theta
\end{split}
\end{equation}
\begin{figure*}[!t]
    \centering
    \includegraphics[width=\textwidth]{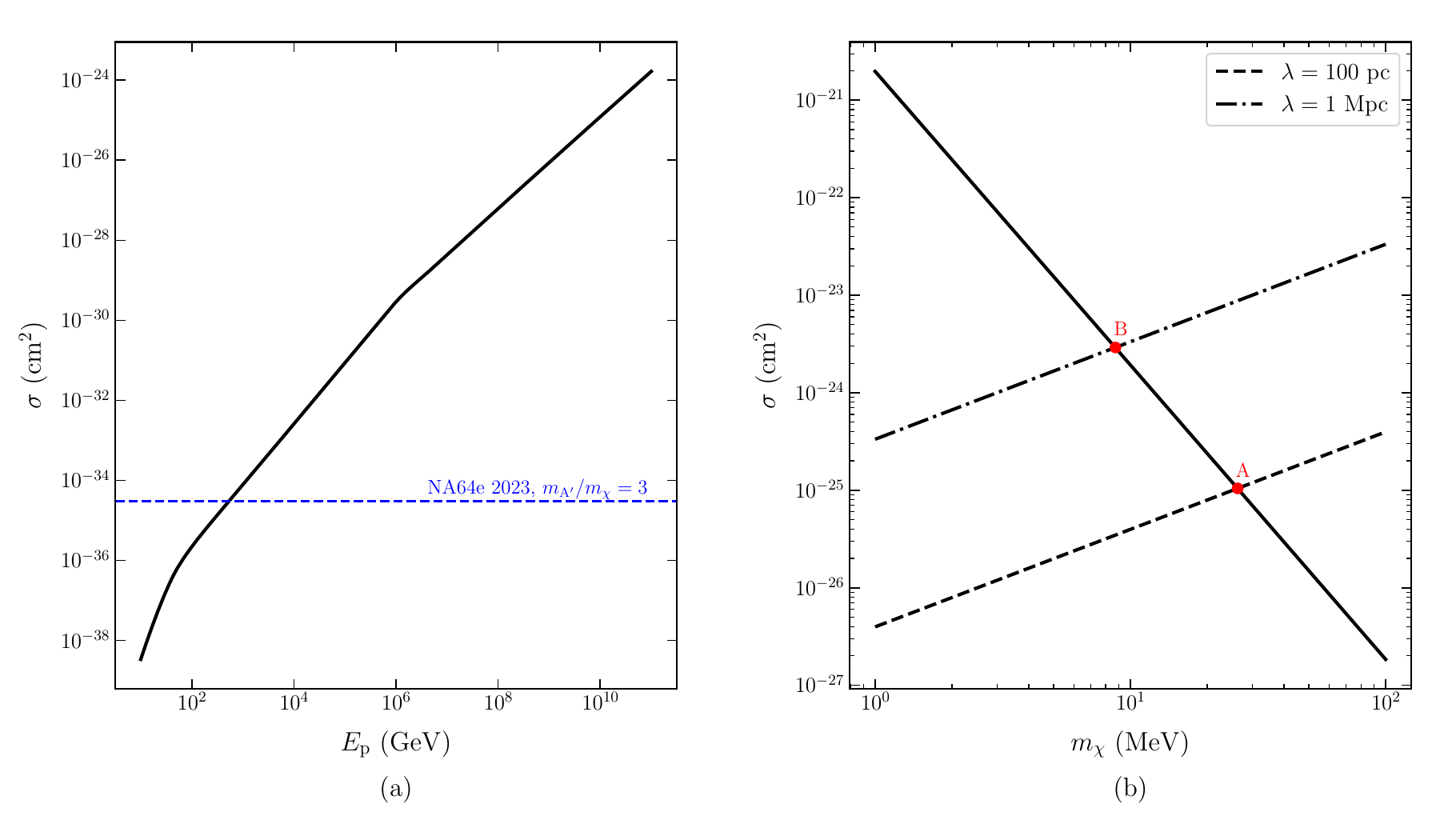}

    \captionsetup{
        width=\textwidth,
        justification=raggedright,
        singlelinecheck=false
    }

    \caption{(a) Scattering cross section as a function of the energy of the incident nucleon
        for a fixed dark matter mass $m_{\mathrm{DM}} = 25~\mathrm{MeV}$. The upper limit from NA64 (2023) \cite{Andreev:2023, Gninenko:2023} ($m_{\mathrm{DM}} \sim 25~\mathrm{MeV}$) are shown by the blue dashed line        
        (b) Scattering cross section as a function of the dark matter mass
        for a fixed energy of the incident proton $E_{\rm p} = 10^{19}~\mathrm{eV}$. The dashed lines show the estimates presented in Fig. 1, and the intersection points are highlighted.}
    \label{fig:sigma_energy_mass}
\end{figure*}
where $x = Q^2 / (-2p\cdot q)$, the squared momentum transfer is given by  $Q^2 = -q^2  = 2m_{\chi}T_{\chi}$ and $-2p\cdot q = 2E_{\rm p}T_{\chi} - 2\lvert \vec{p}\rvert\lvert \vec{k'}\rvert \cos \theta  $, $T_{\chi}$ is a kinetic energy of the final dark-matter particle, $E_{\rm p}$ is an energy of the incident proton, $q = p - p'$. All kinematic quantities are defined in the rest frame of the dark-matter particle. The upper limit $\cos\theta_{\rm max}$ is determined by the condition $x < 1$. The formulas are derived in the deep inelastic scattering (DIS) within the framework of the parton model. The expression for the matrix element is:

\begin{equation}
\begin{split}
\overline{\lvert \mathcal{M}(x)\rvert^2} 
&= \frac{g^2_{\chi} g^2_{\rm q}}{(Q^2 + m^2_{\rm A'})^2} \big[ 16 (x m_{\chi} E_{\rm p})^2 - 8 x m_{\chi} E_p Q^2 \\
&\quad - 4 Q^2 m^2_{\chi} - 4 Q^2 x^2 m^2_{\rm p} + 2 (Q^2)^2 \big]
\end{split}
\end{equation}

The parton distribution functions $f_{\rm q}(x, Q^2)$ from up and down quarks are taken from the results of CT14lo as implemented in the LHAPDF package \cite{Dulat:2015mca, Buckley:2015}. However, in the course of the calculation, the Bjorken variable $x$ reaches values as small as $10^{-13}$, whereas the CT14 parton distribution functions are only defined down to $x=10^{-9}$. Therefore, at high energies ($E_{\rm p} \gtrsim 10^6 \, \mathrm{GeV} $), we employ the quark distributions obtained from a Froissart-bound parametrization \cite{Block:2013qcd}, which allows their extrapolation to ultra-small values of Bjorken \(x\) \cite{Block:2013uhe}. We perform the numerical integration to determine the cross section as a function of the incident-particle energy. At low energies, we use the standard parton distribution functions, while at high energies we employ the logarithmic parametrization of \cite{Block:2013qcd}, smoothly interpolating between the two regimes using a second-order smoothing function.

We then calculate the cross section as a function of the dark-matter particle mass at a fixed energy of $E_{\rm p}=10^{19} \,\mathrm{eV}$. In this case, we use exclusively the logarithmic parton distributions obtained from the Froissart-bound parametrization. The result is shown in Fig.~2.

As can be seen, at low energies the cross section is consistent with existing experimental constraints \cite{Andreev:2023, Gninenko:2023}, while at high energies ($E_{\rm p} \sim 10^{19}- 10^{20}~ \mathrm{eV}$) it reaches sufficiently large values ($\sigma \sim 10^{-25} - 10^{-24}\,\mathrm{cm}^2 $) required for scattering of nucleons off dark matter (panel (a) of Fig.~2). 

Figure~2(b) shows that the cross section depends rather strongly on the dark-matter particle mass: 
as the mass increases, the cross section decreases. 
In contrast, in Fig.~1 the required cross section was found to increase with the mass. 
The relevant region of parameter space lies approximately at the intersection of these curves and is therefore quite narrow, 
corresponding to a mass of order $25~\mathrm{MeV}$ and a cross section in the range 
$10^{-25}$--$10^{-24}~\mathrm{cm}^2$. 
In this work, we consider the most likely scenario to be nucleon scattering off dark matter occurring in the Galactic core, 
with a characteristic scale $\lambda \sim 100~\mathrm{pc}$ (A on Fig.~2(b)). In the scenario where the secondary process takes place within a large-scale structure filament (B on Fig.~2(b)), a detailed evaluation accounting for the density profile has not been performed.

\section{Conclusions}
\label{sec:5}
We showed that in some simple dark-matter models with dark photons, the scattering cross sections between dark matter and nucleons attained large values in the high-energy regime. A scenario was considered in which initially non-relativistic dark matter underwent a collision with an incoming ultra-high-energy nucleon in the nucleus of a galaxy, after which it effectively retransmitted the acquired kinetic energy and transferred it to a baryon close to the observer. 

Our analysis has been restricted to the simplest example, and the required growth of the cross sections may also be realized in other models. This suggests that dark matter might be searched for in interactions at ultra-high energies, for example in processes involving cosmic rays.

Due to the inhomogeneous distribution of gas in a galaxy, the expected signal would exhibit a significant anisotropy if the second process goes in the Milky Way, this in poor agreement with the results of the High Resolution Fly’s Eye (HiRes) experiment. If, instead, one considers a scenario in which the interaction between dark matter and gas occurs in a large-scale structure filament, the picture changes, and the anisotropy becomes consistent with observations \cite{Troitsky:2021}.
\label{sec:concl}

\section*{Acknowledgments}
This work is supported in the framework of the State project "Science" by the Ministry of Science and Higher Education of the Russian Federation under the contract 075-15-2024-541.

The author thanks S. Troitsky for proposing the original idea and for numerous discussions of this work, as well as D. Kirpichnikov and E. Krukova for valuable comments and useful insights.
\bibliography{dark}
\end{document}